\documentclass[10pt,conference]{IEEEtran}

\usepackage{amsmath, amsfonts}
\usepackage{algorithm, algorithmic}
\usepackage{graphicx, comment}

\def\bigorder{{\mathcal{O}}}

\def\NN{{\mathbb{N}}}

\def\EE{{\mathsf{E}}}
\def\LB{{\mathrm{LB}}}
\def\UB{{\mathrm{UB}}}

\def \Theorem#1{{\it Theorem~#1}}

\def \Corollaries#1{{\it Corollaries~#1}}

\def \Figure#1{{\rm Fig.~#1}}

\def \Table#1{{\rm TABLE~#1}}

\newtheorem{theorem}{Theorem}

\newtheorem{corollary}{Corollary}
\def\Algorithm#1{{\rm Algorithm~#1}}

\newenvironment{thmproof}[1]
{\noindent\hspace{2em}{\it #1 }}
{\hspace*{\fill}~\QED\par\endtrivlist\unskip}

\def \markalgline#1{\newcounter{ctrline#1} \setcounter{ctrline#1}{\value{ALC@line}}}
\def \readalgline#1{\arabic{ctrline#1}}

\def\picchk{\put(0,0){\circle{30}}\put(0,0){\makebox(0,0){$+ $}}}

\def\picvar{\put(0,0){\circle{30}}\put(0,0){\makebox(0,0){$\bullet$}}}

\begin{document}

\title{Upper Bounding the Performance of Arbitrary Finite LDPC Codes on Binary Erasure Channels}

\author{\authorblockN{Chih-Chun Wang\thanks{This research was supported in part by the Army Research
Laboratory Collaborative Technology Alliance under Contract No.\
DAAD 19-01-2-0011.}}
\authorblockA{School of Electrical \& Computer Engineering\\
Purdue University\\
West Lafayette, IN 47907, USA\\
Email: chihw@purdue.edu} \and
\authorblockN{Sanjeev R.\ Kulkarni}
\authorblockA{Dept.\ of Electrical Engineering\\
Princeton University\\
Princeton, NJ 08544, USA\\
Email: kulkarni@princeton.edu} \and
\authorblockN{H.\ Vincent Poor}
\authorblockA{Dept.\ of Electrical Engineering\\
Princeton University\\
Princeton, NJ 08544, USA\\
Email: poor@princeton.edu}
 }
%

\maketitle

\begin{abstract}
Assuming iterative decoding for binary erasure channels (BECs), a
novel tree-based technique for upper bounding the bit error rates
(BERs) of arbitrary, finite low-density parity-check (LDPC) codes
is provided and the resulting bound can be evaluated for all
operating erasure probabilities, including both the waterfall and
the error floor regions. This upper bound can also be viewed as a
narrowing search of stopping sets, which is an approach different
from the stopping set enumeration used for lower bounding the
error floor. When combined with optimal leaf-finding modules, this
upper bound is guaranteed to be tight in terms of the asymptotic
order. The Boolean framework proposed herein further admits a
composite search for even tighter results.
For comparison, a refinement of the algorithm is capable of
exhausting all stopping sets of size $\leq 13$ for irregular LDPC
codes of length $n\approx500$, which requires ${500\choose
13}\approx1.67\times 10^{25}$ trials if a brute force approach is
taken. These experiments indicate that this upper bound can be
used both as an analytical tool and as a deterministic
worst-performance (error floor) guarantee, the latter of which is
crucial to optimizing LDPC codes for extremely low BER
applications, e.g., optical/satellite communications.

\end{abstract}

\section{Introduction}

The bit error rate (BER) curve of any {\it fixed}, {\it finite},
low-density parity-check (LDPC) code on binary erasure channels
(BECs) is completely determined by its stopping set distribution.
Due to the prohibitive cost of computing the entire stopping set
distribution \cite{YedidaSudderthBouchaud01}, in practice, the
waterfall threshold of the BER is generally approximated by the
density evolution and pinpointed by the Monte-Carlo simulation,
while the error floor is lower bounded by semi-exhaustively
identifying the dominant stopping sets \cite{Richardson03} or by
importance sampling \cite{HolzlohnerMahadevanMenyukMorrisZweck05}.
Even computing the size of the minimum stopping sets has been
proved to be an NP-hard problem \cite{KrishnanShankar0000}, which
further shows the difficulty of constructing the entire stopping
set distribution. Other research directions related to the finite
code performance include
\cite{AmraouiUrbankeMontanariRichardson04}
and~\cite{DiProiettiTelatarRichardsonUrbanke02} on the average
performance of finite {\it code ensembles} on BECs and its scaling
law, and a physics-based asymptotic approximation for Gaussian
channels \cite{StepanovChernyakChertkovVasic0000}.


In this paper, only BECs will be considered. We focus on upper
bounding the BER curves of arbitrary, fixed, finite parity check
codes under iterative decoding, and the frame error rate (FER)
will be treated as a special case. Experiments are conducted for
the cases $n=24,50,72, 144$, which demonstrate the superior
efficiency of the proposed algorithm. Application of this bound to
 finite code optimization is deferred to a companion paper.


\section{Boolean Expressions with Nested Structures}
Without loss of generality, we assume the all-zero codeword is
transmitted for notational simplicity.

For BECs, a decoding algorithm for bit $x_i\in\{0,e\}$,
$i\in[1,n]$ is equivalent to a function
$g_{i}:\{0,e\}^n\mapsto\{0,e\}$, where ``$e$" represents an erased
bit and $n$ is the codeword length. In this paper, $f_{i,l}$,
$\forall i\in[1,n]$, is used to denote the iterative decoder for
bit $x_i$ after $l$ iterations. If we further rename the element
``$e$" by ``$1$," $f_{i,l}$ becomes a Boolean function, and the
BER for bit $x_i$ after $l$ iterations is simply
$p_{i,l}=\EE\{f_{i,l}\}$. Another advantage of this conversion is
that the decoding operation at the variable node then becomes
``$\cdot$", the binary {\sf AND} operation, and the operation at
the parity check node becomes ``$+$", the binary {\sf OR}
operation.
\begin{figure}
\centering \setlength{\unitlength}{.085mm}
    \begin{picture}(500,180)(0,30)

    \put(10,12){$i=$}
    \put(82, 80){
    \put(0,-18){\circle{36}}
    \put(-12,-68){1}
    }
    \put(162, 80){
    \put(0,-18){\circle{36}}
    \put(-12,-68){2}
    }
    \put(242, 80){
    \put(0,-18){\circle{36}}
    \put(-12,-68){3}
    }
    \put(322, 80){
    \put(0,-18){\circle{36}}
    \put(-12,-68){4}
    }
    \put(402, 80){
    \put(0,-18){\circle{36}}
    \put(-12,-68){5}
    }
    \put(482, 80){
    \put(0,-18){\circle{36}}
    \put(-12,-68){6}
    }
%
    \put(162, 160){
    \put(-16,0){\framebox(32,32){}}
    \put(-12,42){1}
    }
%
    \put(242, 160){
    \put(-16,0){\framebox(32,32){}}
    \put(-12,42){2}
    }
%
    \put(322, 160){
    \put(-16,0){\framebox(32,32){}}
    \put(-12,42){3}
    }
%
    \put(402, 160){
    \put(-16,0){\framebox(32,32){}}
    \put(-12,42){4}
    }
\put(92, 202){$j=$}
%
    \put(82,80){\line(1,1){80}}
    \put(162,80){\line(0,1){80}}
    \put(242,80){\line(-1,1){80}}
    \put(162,80){\line(1,1){80}}
    \put(322,80){\line(-1,1){80}}
    \put(482,80){\line(-3,1){240}}
    \put(242,80){\line(1,1){80}}
    \put(322,80){\line(0,1){80}}
    \put(402,80){\line(-1,1){80}}
    \put(82,80){\line(4,1){320}}
    \put(402,80){\line(0,1){80}}
    \put(482,80){\line(-1,1){80}}
\end{picture}
\caption{A simple parity check code.\label{fig:simple-code} }
\end{figure}
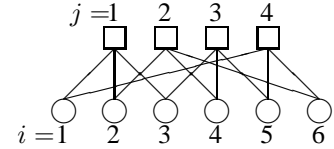

For example, suppose we further use $f_{i\rightarrow j, l}$ to
represent the message from variable node $i$ to check node $j$
during the $l$-th iteration, and consider the simple code
described in \Figure{\ref{fig:simple-code}}. The iterative
decoders $f_{2,l}$, $l\in\{1,2\}$, for bit $x_2$ then become
\begin{eqnarray}
f_{2,1}&=&x_2(x_1+x_3)(x_4+x_6)\nonumber\\
f_{2,2}&=&x_2\left(x_1(f_{5\rightarrow4,1}+f_{6\rightarrow4,1})+x_3(f_{4\rightarrow3,1}+f_{5\rightarrow3,1})\right)\nonumber\\
&&\cdot\left(x_4(f_{3\rightarrow3,1}+f_{5\rightarrow 3,1})+x_6(f_{1\rightarrow 4,1}+f_{5\rightarrow 4,1})\right)\nonumber\\
&=&x_2\left(x_1(x_5+x_6)+x_3(x_{4}+x_{5})\right)\nonumber\\
&&\cdot\left(x_4(x_{3}+x_{5})+x_6(x_{1}+x_{5})\right).\label{eq:original-form}
\end{eqnarray}
 The final decoder of bit $x_2$ is
$f_{2}:=\lim_{l\rightarrow\infty}f_{2,l}$, and in this example,
$f_2=f_{2,2}$. Although (\ref{eq:original-form}) admits a
beautiful nested structure, the repeated appearance of many
Boolean input variables, also known as short ``cycles," poses a
great challenge to the evaluation of the BER $p_{2}=\EE\{f_{2}\}$.
One solution is to first simplify (\ref{eq:original-form}) by
expanding the nested structure into a sum-product Boolean
expression \cite{YedidaSudderthBouchaud01}:
\begin{eqnarray}
f_2=x_1x_2x_4x_5+x_2x_3x_4+x_2x_3x_5x_6.\label(\label{eq:sum-product}
\end{eqnarray}
$\EE\{f_{2}\}$ can then be evaluated by the inclusion-exclusion
principle: $p_{2}=\epsilon^3+2\epsilon^4-2\epsilon^5$, where
$\epsilon$ is the erasure probability.


It can be proved that each product term corresponds to an
irreducible stopping set (IRSS) and vice versa. Instead of
constructing the exact expression of $f_i$, 
if only a small subset of these IRSSs is identified, say
``$x_2x_3x_4$," then a lower bound $\EE\{f_{\LB,2}\}=\epsilon^3$
can be obtained, where
\begin{eqnarray}
f_{\LB,2}=x_2x_3x_4\leq f_2,\mbox{~and~}
\EE\{f_{LB,2}\}=\epsilon^3\leq \EE\{f_2\}.\nonumber
\end{eqnarray}
The major challenge of this approach is to ensure that all IRSSs
of the minimum weight/order are exhausted. Furthermore, even when
all IRSSs of the minimum weight are exhausted, this lower bound is
tight only in the high signal-to-noise ratio (SNR) regime. Whether
the SNR of interest is high enough can only be determined by
Monte-Carlo simulations and by extrapolating the waterfall region.

An upper bound can be constructed by iteratively computing the
sum-product form of $f_{2,l_0+1}$ from that  of $f_{2,l_0}$. To
counteract the exponential
 growth rate of the number of
product terms, during each iteration, we can ``relax" and ``merge"
some of the product terms so that the complexity is kept
manageable \cite{YedidaSudderthBouchaud01}. For example, $f_{2,2}$
in (\ref{eq:sum-product}) can be relaxed and merged as follows.
\begin{eqnarray} f_{2,2}&=&x_1x_2x_4x_5+x_2x_3x_4+x_2x_3x_5x_6\nonumber\\&\leq&
1 x_2x_4 1+x_21x_4+x_2x_3x_5x_6\nonumber\\
&=&x_2x_4+x_2x_3x_5x_6,\nonumber
\end{eqnarray}
so that the number of product terms is reduced to two.
Nonetheless, the minimal number of product terms required to
generate a tight upper bound grows very fast and  good/tight
results were reported only for the cases $n\leq 20$. In contrast,
we construct an efficient upper bound $\UB_i\geq \EE\{f_i\}$ by
preserving much of the nested structure, so that tight upper
bounds can be obtained for $n=100$--$300$. Furthermore, the
tightness of our bound can be verified with ease, which was absent
in the previous approach. Combined with the lower bound
$\EE\{f_{\LB,i}\}$, the finite code performance can be efficiently
bracketed for the first time.


\section{An Upper Bound Based on Tree-Trimming\label{sec:rules}}

Two fundamental observations can be proved as follows.

{\it Observation 1:}  All $f_i$'s  are monotonic functions w.r.t.
all input variables. Namely, $f_i|_{x_j=0}\leq f_i|_{x_j=1}$ for
all $i,j\in[1,n]$, which separates  $f_i$'s from ``arbitrary"
Boolean functions. (Here we use the point-wise ordering such that
$f\leq g$ iff $f({\mathbf{x}})\leq g({\mathbf{x}})$ for all binary
vectors ${\mathbf{x}}$.)

{\it Observation 2:} The correlation coefficient between any pair
of $f_i$ and $f_j$  is always non-negative, i.e., $\EE\{f_i\cdot
f_j\}\geq \EE\{f_i\}\EE\{f_j\}$.

In this section, we assume that $f_v=g\cdot h$ or $f_c=g+h$ for variable or
check node operations respectively. Define ${\mathbf x}_g\subseteq
\{x_1,\cdots, x_n\}$ as the set of input variables upon which the Boolean
function $g$ depends, e.g., if $g=x_1x_2+x_7$, then ${\mathbf
x}_g=\{x_1,x_2,x_7\}$. Similary, we can define ${\mathbf x}_h$.
\subsection{Rule 0}
\begin{quote}
If ${\mathbf x}_g\cap{\mathbf x}_h=\emptyset$, namely, there is no
repeated node in the input arguments, then
\begin{eqnarray}
\EE\{f_v\}&=&\EE\{g\}\EE\{h\}\nonumber\\
\EE\{f_c\}&=&\EE\{g\}+\EE\{h\}-\EE\{g\}\EE\{h\}.\nonumber
\end{eqnarray}
\end{quote}

\subsection{Rule 1 -- A Simple Relaxation}
Suppose ${\mathbf x}_g\cap{\mathbf x}_h\neq \emptyset$, namely,
there are repeated nodes in the input arguments. By Observation~2
and Rule~0, we have
\begin{eqnarray}
\EE\{f_c\}&=&\EE\{g\}+\EE\{h\}-\EE\{g\cdot
h\}\nonumber\\
&\leq&\EE\{g\}+\EE\{h\}-\EE\{g\}\EE\{h\}.\label{eq:indirect-chk}
\end{eqnarray}

\begin{figure}
\begin{tabular}{cc}
\setlength{\unitlength}{.15mm}
    \begin{picture}(160,125)(0,25)

\put(80,135){\picchk}

\put(0,25){\line(2,3){50}}

\put(50,100){\line(3,2){30}}

\put(160,25){\line(-2,3){50}}

\put(110,100){\line(-3,2){30}}

\put(0,25){\line(1,0){160}}

\put(80,55){\line(-2,3){30}}

\put(80,55){\line(2,3){30}}

\put(65,25){\line(0,1){30}}

\put(95,25){\line(0,1){30}}

\put(65,55){\line(1,0){30}}

\put(67,35){$x_{i_0}$}

\end{picture}&
\setlength{\unitlength}{.15mm}
    \begin{picture}(200,125)(0,25)

\put(100,135){\picchk}

\put(0,25){\line(2,3){50}}

\put(50,100){\line(5,2){50}}

\put(200,25){\line(-2,3){50}}

\put(150,100){\line(-5,2){50}}

\put(0,25){\line(1,0){95}}

\put(95,25){\line(0,1){30}}

\put(65,55){\line(1,0){30}}

\put(65,25){\line(0,1){30}}

\put(67,35){$x_{i_0}$}

\put(105,25){\line(1,0){95}}

\put(105,25){\line(0,1){30}}

\put(105,55){\line(1,0){30}}

\put(135,25){\line(0,1){30}}

\put(107,35){$x'_{i_0}$}

\put(80,55){\line(-2,3){30}}

\put(120,55){\line(2,3){30}}

\end{picture}\\
(a) Original & (b) Relaxation
\end{tabular}
\caption{Rule 1: A simple relaxation for check nodes.
\label{fig:indirect-chk}}
\end{figure}
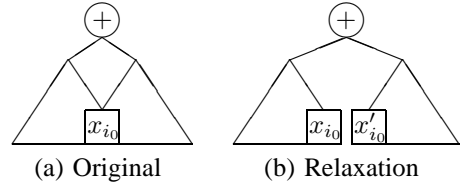

The above rule suggests that when the incoming messages of a check
node are dependent, the error probability of the outgoing message
can be upper bounded by assuming the incoming ones are
independent. Furthermore, Rule~1 does not change the order of
error probability, as can be seen in (\ref{eq:indirect-chk}), but
only modifies the multiplicity term. Due to the random-like
interconnection within the code graph, for most cases, $g$ and $h$
are ``nearly independent" and the multiplicity loss is not
significant. The realization of Rule~1 is illustrated in
\Figure{\ref{fig:indirect-chk}}, in which we assume that $x_{i_0}$
is the repeated node.

\subsection{Rule 2 -- The Pivoting Rule}
\begin{figure}
\begin{tabular}{cc}
\setlength{\unitlength}{.15mm}
    \begin{picture}(160,125)(0,25)

\put(80,135){\picvar}

\put(0,25){\line(2,3){50}}

\put(50,100){\line(3,2){30}}

\put(160,25){\line(-2,3){50}}

\put(110,100){\line(-3,2){30}}

\put(0,25){\line(1,0){160}}

\put(80,55){\line(-2,3){30}}

\put(80,55){\line(2,3){30}}

\put(65,25){\line(0,1){30}}

\put(95,25){\line(0,1){30}}

\put(65,55){\line(1,0){30}}

\put(67,35){$x_{i_0}$}

\end{picture}&
\setlength{\unitlength}{.15mm}
    \begin{picture}(350,188.5)(0,25)

\put(0,0){\put(85,132.5){\picvar}

\put(0,25){\line(2,3){50}}

\put(50,100){\line(2,1){35}}

\put(170,25){\line(-2,3){50}}

\put(120,100){\line(-2,1){35}}

\put(0,25){\line(1,0){65}}

\put(105,25){\line(1,0){65}}

\put(80,55){\line(-2,3){30}}

\put(90,55){\line(2,3){30}}

\put(65,25){\line(0,1){30}}

\put(105,25){\line(0,1){30}}

\put(65,55){\line(1,0){15}}

\put(90,55){\line(1,0){15}}

\put(80,30){$1$}}

\put(40,132.5){\vector(1,0){30}}

\put(10,117.5){\framebox(30,30){$x_{i_0}$}}

\put(180,0){\put(85,132.5){\picvar}

\put(0,25){\line(2,3){50}}

\put(50,100){\line(2,1){35}}

\put(170,25){\line(-2,3){50}}

\put(120,100){\line(-2,1){35}}

\put(0,25){\line(1,0){65}}

\put(105,25){\line(1,0){65}}

\put(80,55){\line(-2,3){30}}

\put(90,55){\line(2,3){30}}

\put(65,25){\line(0,1){30}}

\put(105,25){\line(0,1){30}}

\put(65,55){\line(1,0){15}}

\put(90,55){\line(1,0){15}}

\put(80,30){$0$}}

\put(175,198.5){\picchk}

\put(85,147.5){\line(5,2){90}}

\put(265,147.5){\line(-5,2){90}}

\end{picture}\\
(a) Original & (b) Decoupled
\end{tabular}
\caption{Rule 2: A pivoting rule for variable nodes.
\label{fig:pivoting-rule}}
\end{figure}
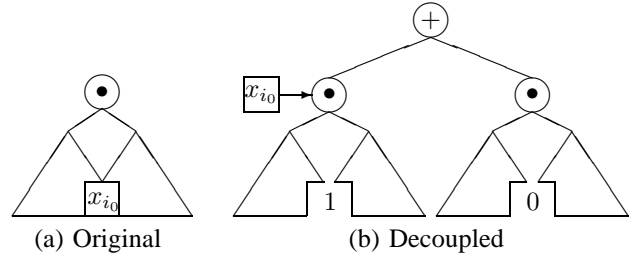

Consider the simplest case in which ${\mathbf x}_g\cap{\mathbf
x}_h=\{x_{i_0}\}$. By Observation~1, we have
\begin{eqnarray}
f_v&=&g_v|_{x_{i_0}=0}\cdot h_v|_{x_{i_0}=0}\nonumber\\
&&+x_{i_0}\cdot g_v|_{x_{i_0}=1}\cdot
h_v|_{x_{i_0}=1}.\label{eq:pivoting}
\end{eqnarray}
The realization of the above equation is demonstrated in
\Figure{\ref{fig:pivoting-rule}}.
Once the tree in \Figure{\ref{fig:pivoting-rule}(a)} is
transformed to \Figure{\ref{fig:pivoting-rule}(b)}, all messages
entering the variable nodes become independent since
$g_v|_{x_{i_0}=0}$ and $h_v|_{x_{i_0}=0}$ then share no repeated
input variables.
By reapplying Rules~0 and~1, the output $\EE\{f_v\}$ is upper
bounded by
\begin{eqnarray}
\EE\{f_v\}&\leq&
\EE\{f_v|x_{i_0}=0\}+\EE\{x_{i_0}\}\EE\{f_v|x_{i_0}=1\}\nonumber\\
&&-
\EE\{f_v|x_{i_0}=0\}\EE\{x_{i_0}\}\EE\{f_v|x_{i_0}=1\},\nonumber
\end{eqnarray}
where for all $b\in\{0,1\}$,
\begin{eqnarray}
\EE\{f_v|x_{i_0}=b\}=\EE\{g_v|x_{i_0}=b\}
\EE\{h_v|x_{i_0}=b\}.\nonumber
\end{eqnarray}

Note: the pivoting rule (\ref{eq:pivoting}) does not incur any
performance loss. The actual loss during this step is the
multiplicity loss resulted from reapplying Rule~1. Therefore,
Rule~2 preserves the asymptotic order of $\EE\{f_v\}$ as does
Rule~1.

\subsection{The Algorithm}
Rules~0 to~2 are designed to upper bound the expectation of single
operations with zero or a single overlapped node. Once carefully
concatenated, they can be used to construct $\UB_{i}$ for the
infinite tree with many repeated nodes, while preserving most of
the nested structure.
\begin{algorithm}[t]\footnotesize
 \caption{A tree-based method to upper bound $p_i$.\label{alg:tree1}}

{\bf Initialization:} Let $\mathcal T$ be a tree containing only
 the target $x_{i}$ variable node.

\vspace{-.2cm}

\hrulefill

\vspace{-.1cm}

\begin{algorithmic}[1]
\REPEAT

\STATE Find the next leaf variable node, say $x_j$.
\markalgline{line:find-leaf}

\IF{there exists another non-leaf $x_j$ variable node in $\mathcal
T$}

    \IF{the youngest common ancestor of the leaf $x_j$ and any other
existing non-leaf $x_j$, denoted as ${\mathsf{yca}}(x_j)$, is a
check node}

%
%

            \STATE As suggested by Rule~1, the new leaf node $x_j$ can be
directly included as if there are no other $x_j$'s  in $\mathcal
T$.


    \ELSIF{${\mathsf{yca}}(x_j)$ is a variable node}

%
%

            \STATE As suggested by Rule~2, a pivoting construction involving
tree duplication is initiated, which is illustrated in
\Figure{\ref{fig:pivoting-rule}}.


    \ENDIF

\ENDIF

\STATE Construct  the immediate check node children and variable
node grand children of $x_j$ as in the support tree of the
corresponding Tanner graph.

\UNTIL{the size of $\mathcal T$ exceeds the preset limit.}

\STATE Hardwire all remaining leaf nodes to $1$.

 \STATE $\UB_{i}$ is evaluated by invoking Rules~0 and~1. Namely,
all incoming edges are blindly assumed to be independent.
\markalgline{line:eval}
\end{algorithmic}
\end{algorithm}
\begin{theorem}
The concatenation in \Algorithm{\ref{alg:tree1}} is guaranteed to
find an upper bound $\UB_{i}$ for $p_{i}$ of the infinite
tree.\label{thm:upper-boundedness}
\end{theorem}

The proof of \Theorem{\ref{thm:upper-boundedness}} involves
 the graph theoretic properties of
${\mathsf{yca}}(x_j)$ and an incremental tree-revealing argument.
Some other properties of \Algorithm{\ref{alg:tree1}} are listed as
follows.
\begin{itemize}
\item The only computationally expensive step is when Rule~2 is
invoked, which, in the worst case, may double the tree size and
thus reduces the efficiency of this algorithm.

\item Rule~1, being the only relaxation rule, saves much computational cost
by ignoring repeated nodes.

\item Once the tree construction is completed, evaluating
$\UB_{i}$ for any $\epsilon\in[0,1]$ is efficient with complexity
$\bigorder\left(|{\mathcal{T}}|\log(|{\mathcal{T}}|)\right)$,
where $|{\mathcal{T}}|$ is the size of $\mathcal T$.

\item The preset size limit of $\mathcal T$ provides a tradeoff
between computational resources and the tightness of the resulting
$\UB_{i}$. 
One can terminate the program early before the tightest results
are obtained, as long as the intermediate results have  met the
evaluation/design requirements.
\end{itemize}

This tree-based approach corresponds to a narrowing search of
stopping sets. By denoting $f_{{\mathcal T },t}$ as the
corresponding Boolean function of the
tree\footnote{\Algorithm{\ref{alg:tree1}} consists of the tree
construction stage and the upper bound computing stage
(Line~\readalgline{line:eval}). In this paper, $f_{{\mathcal T
},t}:\{0,1\}^n\mapsto \{0,1\}$ is defined on the constructed tree,
which will then be used on evaluating $\UB_{i}$. } $\mathcal T$ at
time $t$, we have
\begin{theorem}[A Narrowing Search]
Let
\begin{eqnarray}
{\mathbf{X}}_t:=\{(x_1,\cdots,x_n)\in\{0,1\}^n:
f_{{\mathcal{T}},t}(x_1,\cdots,x_n)=1\}.\nonumber
\end{eqnarray}
We then have
\begin{eqnarray}
\{{\mathrm{all~stopping~sets}}\}\subseteq
{\mathbf{X}}_{t+1}\subseteq {\mathbf{X}}_t, \forall
t\in\NN.\nonumber
\end{eqnarray}\label{thm:narrowing-search}
\end{theorem}

\section{Performance and Related Topics}
\subsection{The Leaf-Finding Module\label{subsec-LF}}
The tightness of $\UB_{i}$ in \Algorithm{\ref{alg:tree1}} depends
heavily on the leaf-finding (LF) module invoked in
Line~\readalgline{line:find-leaf}.
A properly designed LF module is capable of
increasing the asymptotic order of $\UB_i$ by +1 to +3. The
ultimate benefit of an optimal LF module is stated in the
following theorem.
\begin{theorem}[The Optimal LF Module]\label{thm:LF-optimality}
Following the notation in \Theorem{\ref{thm:narrowing-search}},
with an ``optimal" LF module, we have
\begin{eqnarray}
\{{\mathrm{all~stopping~sets}}\}=
\lim_{t\rightarrow\infty}{\mathbf{X}}_{t}.\nonumber
\end{eqnarray}
\end{theorem}

\begin{corollary}[Order Tightness of \Algorithm{\ref{alg:tree1}}]
When combined with an optimal LF module, the $\UB_{i}$ computed by
\Algorithm~\ref{alg:tree1} is tight in terms of the asymptotic
order. Namely, $\exists C>0$ such that
$\frac{\UB_{i}(\epsilon)}{p_{i}(\epsilon)}<C$ for all
$\epsilon\in(0,1]$.
\end{corollary}

In \cite{KrishnanShankar0000}, determining whether a fixed LDPC
codes contains a stopping set of size $\leq t$ is proved to be
NP-hard. By \Theorem{\ref{thm:narrowing-search}}, a
straightforward choice of the LF module, and the complexity
analysis of \Algorithm{\ref{alg:tree1}}, we have
\begin{theorem} Deciding whether the stopping distance is $\leq t$
is fixed-parameter tractable when $t$ is fixed.
\end{theorem}

For all our experiments, an efficient approximation of the optimal
LF module, motivated by the proof of
\Theorem{\ref{thm:LF-optimality}}, is adopted. With reasonable
computational resources, \Algorithm{\ref{alg:tree1}} is capable of
constructing asymptotically tight UBs for LDPC codes of
$n\leq100$. A composite approach will be introduced later, which
further extends the application range to $n\leq300$.

\begin{table*}
\centering
\begin{tabular}{ccc}

(a) $n=50$ & (b) $n=72$& (c) $n=144$\\ \begin{tabular}{lccccc}
\hline \hline
 Order&  3&4& 5&6&7\\
 \hline
Num.\ bits& 3&11&10& 20& 6\\
order*& & &3& 8& 5\\
~~+ multi*&3 &11 & 7 & 12 &  1\\
\hline \hline
\end{tabular}&\begin{tabular}{lcccccc}
\hline \hline
 Order& 2& 4& 5&6&7&8\\
 \hline
Num.\ bits& 4&4&5&28&28&3\\
order*&  &  & 1&11&26&1\\
~~+ multi*&4 &4 &4&17& 2 &\\
\hline \hline
\end{tabular}&\begin{tabular}{lccccc}
\hline \hline
 Order&  2&5& 7&8&9\\
 \hline
order*&4 &3 &7& 27& 2\\
order$>$& & & 6 & 90& 5\\
\hline \hline
\end{tabular}
\end{tabular}
\caption{Performance Statistics: {\rm\scriptsize``{\rm Num.\
bits}" is the number of bits with the specified asymptotic order.
``{\rm order*}" is the num.\ bits with UBs tight only in the
order. ``{\rm + multi*}" is the num.\ bits with UBs tight both in
 the order and in the multiplicity. ``{\rm order $>$}" is the num.\
bits with a $\UB$ of the specified order while no bracketing lower
bound can be established.}\label{tab:perf-stat}}
\end{table*}
\subsection{Confirming the Tightness of $\UB_i$}
To this end, after each time $t$, we first exhaustively enumerate
the elements of minimal weight in ${\mathbf{X}}_t$ and denote the
collection of them as ${\mathbf{X}}_{min}$.
\begin{corollary}[Tightness Confirmation]
If  $\exists {\mathbf{x}}\in{\mathbf{X}}_{min}$ that is a stopping
set, then $\UB_{i}$ is tight in terms of the asymptotic
order.\label{cor:tight-confirm}
\end{corollary}
\begin{corollary}[The Tight Upper and Lower Bound Pair]
Let ${\mathbf{X}}_{min,SS}\subseteq {\mathbf{X}}_{min}$ denote the
collection of all elements ${\mathbf{x}}\in{\mathbf{X}}_{min}$
that are also stopping sets. Then ${\mathbf{X}}_{min,SS}$ exhausts
the stopping sets of the minimal weight, and can be used to derive
a lower bound $\EE\{f_{\LB,i}\}$ that is tight in both the
asymptotic order and the multiplicity.\label{cor:tight-LB} This
exhaustive lower bound 
was not found in any existing papers.
\end{corollary}

\subsection{BER vs.\ FER}
The above discussion has focused on providing $\UB_i$ for a
pre-selected target bit $x_i$. Bounds for the average BER can be
easily obtained by taking averages over bounds for individual
bits. An equally interesting problem is bounding the FER, which
can be converted to the BER as follows. Introduce an auxiliary
variable and check node pair $(x_0,y_0)$, such that the the new
variable node $x_0$ is punctured and the new check node $y_0$ is
connected to all $n+1$ variable nodes from $x_0$ to $x_n$. The FER
of the original code now equals the BER $p_0$ of variable node
$x_0$ and can be bounded by \Algorithm{\ref{alg:tree1}}. Since the
FER depends only on the worst bit performance, it is easier to
construct tight $\UB$ for the FER than for the BER. On the other
hand, $\UB_i$ provides detailed performance prediction for each
individual bit, which is of great use during code analysis.

\subsection{A Composite Approach}

The expectation $\EE\{f_{i}\}$ can be further decomposed as
\begin{eqnarray}
\EE\{f_{i}\}=\sum_{j=1}^{M}\EE\{{\mathcal
A}_j\}\EE\{f_{i}|{\mathcal A}_j\},\nonumber
\end{eqnarray}
where ${\mathcal A}_j$'s are $M$  events partitioning the sample
space. For example, we can define a collection of non-uniform
${\mathcal A}_j$'s by
\begin{eqnarray}
{\mathcal A}_1&=&\{x_0=0\}\nonumber\\
{\mathcal A}_2&=&\{x_0=1, x_7=0\}\nonumber\\
{\mathcal A}_3&=&\{x_0=1, x_7=1\}.\nonumber
\end{eqnarray}
Since for any $j$, $f_{i}|_{{\mathcal A}_{j}}$ is simply another
finite code with a modified Tanner graph,
\Algorithm{\ref{alg:tree1}} can be applied to each
$f_{i}|_{A_{j}}$ respectively and different
$\UB_{i,j}\geq\EE\{f_{i}|_{A_{j}}\}$ will be obtained. A composite
upper bound is now constructed by
\begin{eqnarray}
\mbox{C-UB}_{i}=\sum_{j=1}^{M}\EE\{{\mathcal
A}_j\}\UB_{i,j}\geq\EE\{f_i\}=p_i.\nonumber
\end{eqnarray}
In general, C-UB is able to produce bounds that are +1 or +2 in
the asymptotic order and pushes the application range to
$n\leq300$. The efficiency of C-UB relies on the design of the
non-uniform partition $\{A_j\}$.


\def\hei{2.5cm}
\def\wid{5.5cm}
\def\mhe{-.4cm}

\begin{figure*}
\begin{tabular}{ccc}
\parbox[\hei]{5.5cm}{\vspace{\mhe}\includegraphics[width=6cm,
keepaspectratio=true]{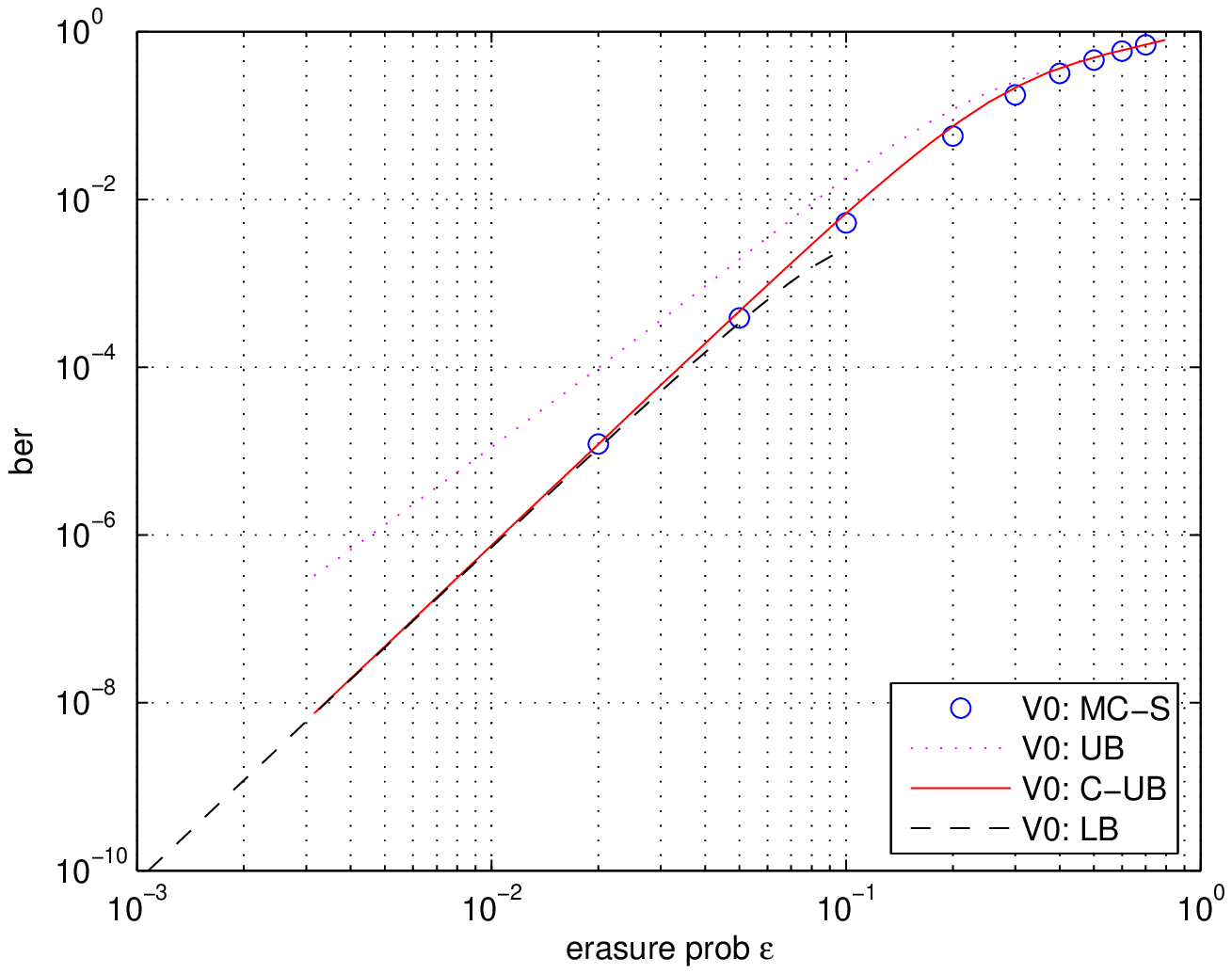}}&\parbox[\hei]{5.5cm}{\vspace{\mhe}\includegraphics[width=6cm,
keepaspectratio=true]{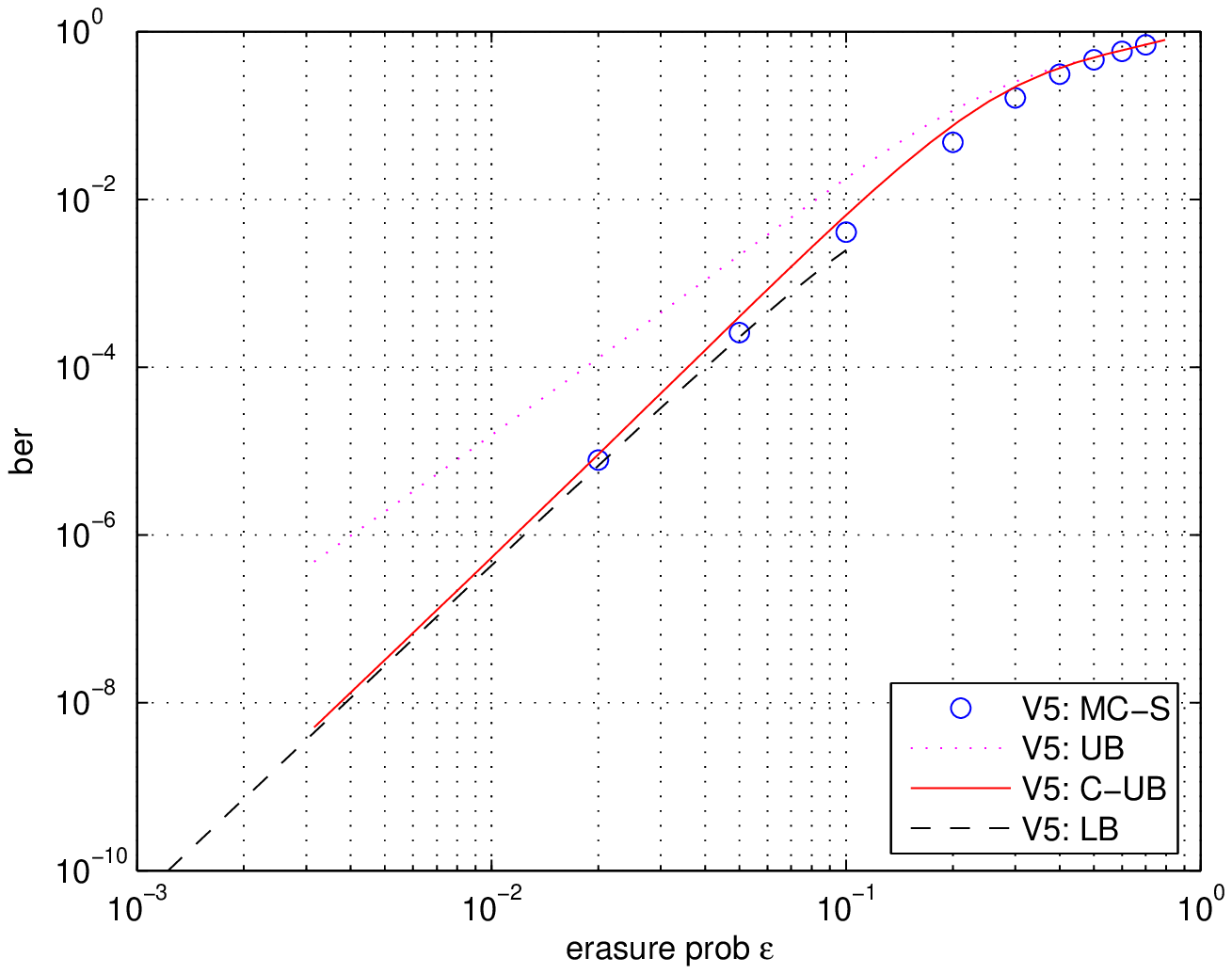}}
&\parbox[\hei]{5.5cm}{\vspace{\mhe}\includegraphics[width=6cm,
keepaspectratio=true]{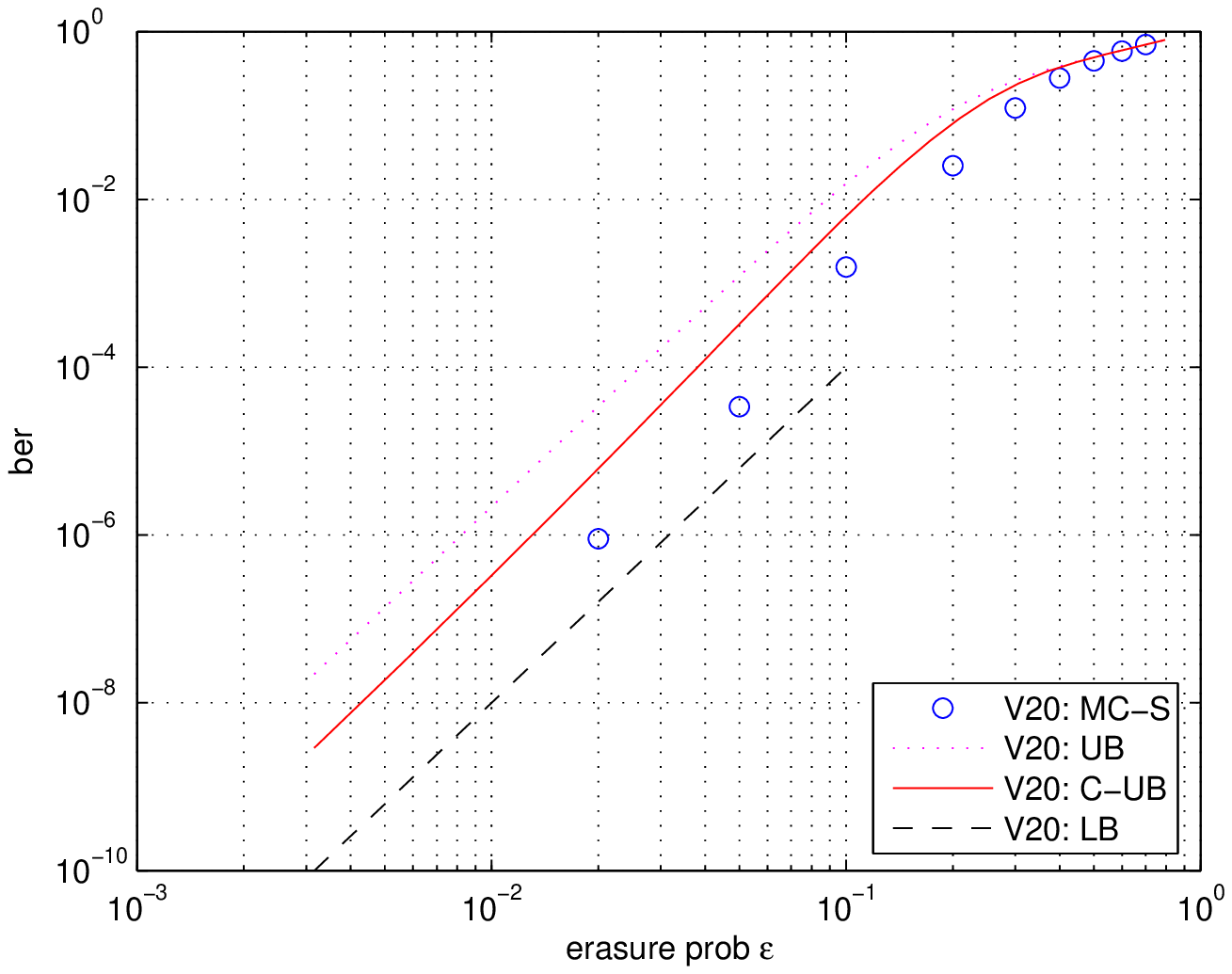}}
\end{tabular}
\caption{Comparisons among the upper bound (UB), the composite
upper bound (C-UB), the Monte-Carlo  simulation (MC-S), and the
side product tight lower bound (LB) for bits~0, 5, and~20 of the
(23,12) binary Golay code. The corresponding asymptotic (order,
multiplicity) pairs obtained by UB, C-UB, and the actual BER is
are V0: $\{(3,10), (4,75), (4,75)\}$, V5: $\{(3,15), (4,50),
(4,45)\}$, and V20: $\{(4,221), (4,27), (4,1)\}$.
\label{fig:Golay}}
\end{figure*}
\begin{figure*}
\begin{tabular}{ccc}
\parbox[\hei]{5.5cm}{\vspace{\mhe}\includegraphics[width=6cm,
keepaspectratio=true]{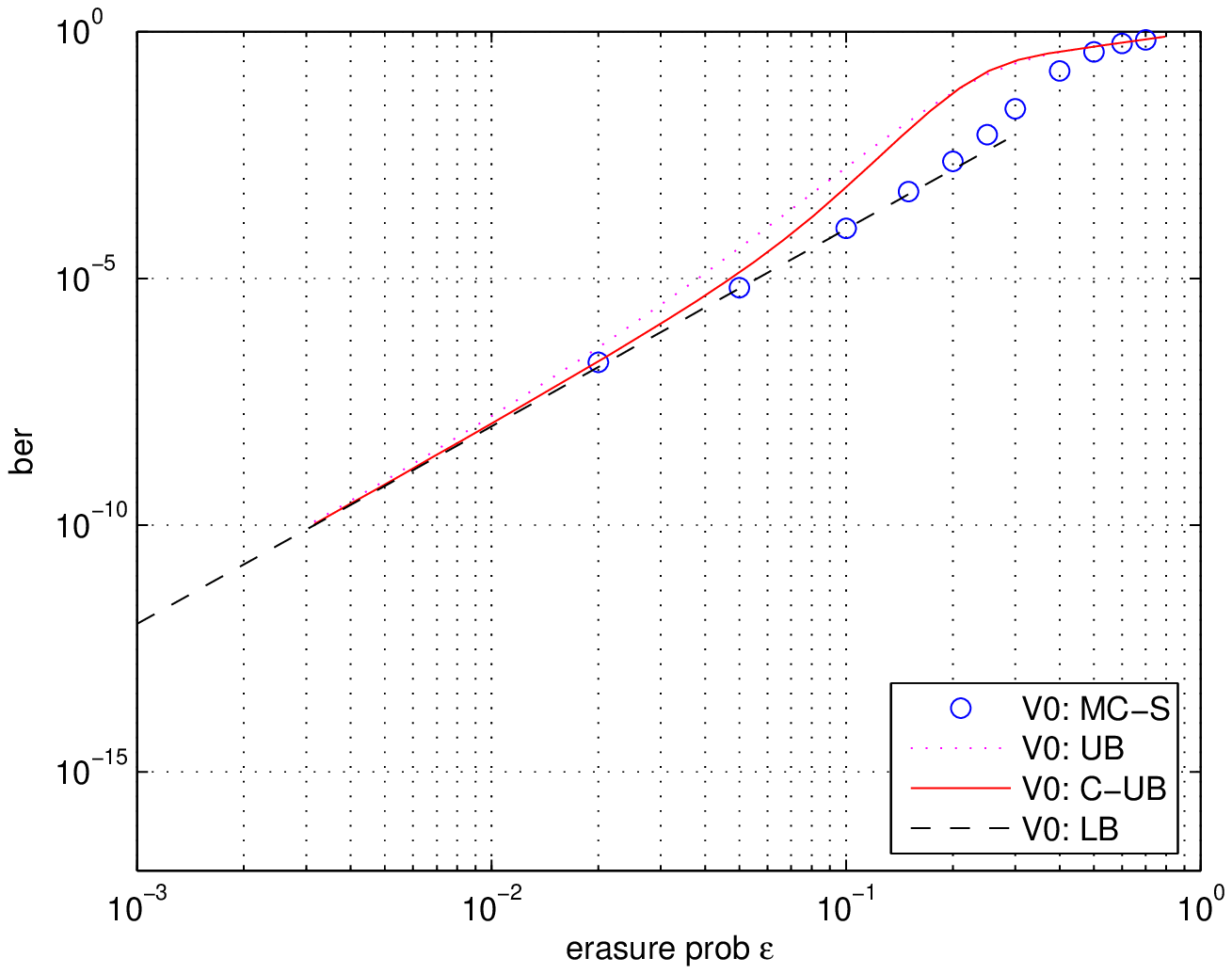}}&\parbox[\hei]{5.5cm}{\vspace{\mhe}\includegraphics[width=6cm,
keepaspectratio=true]{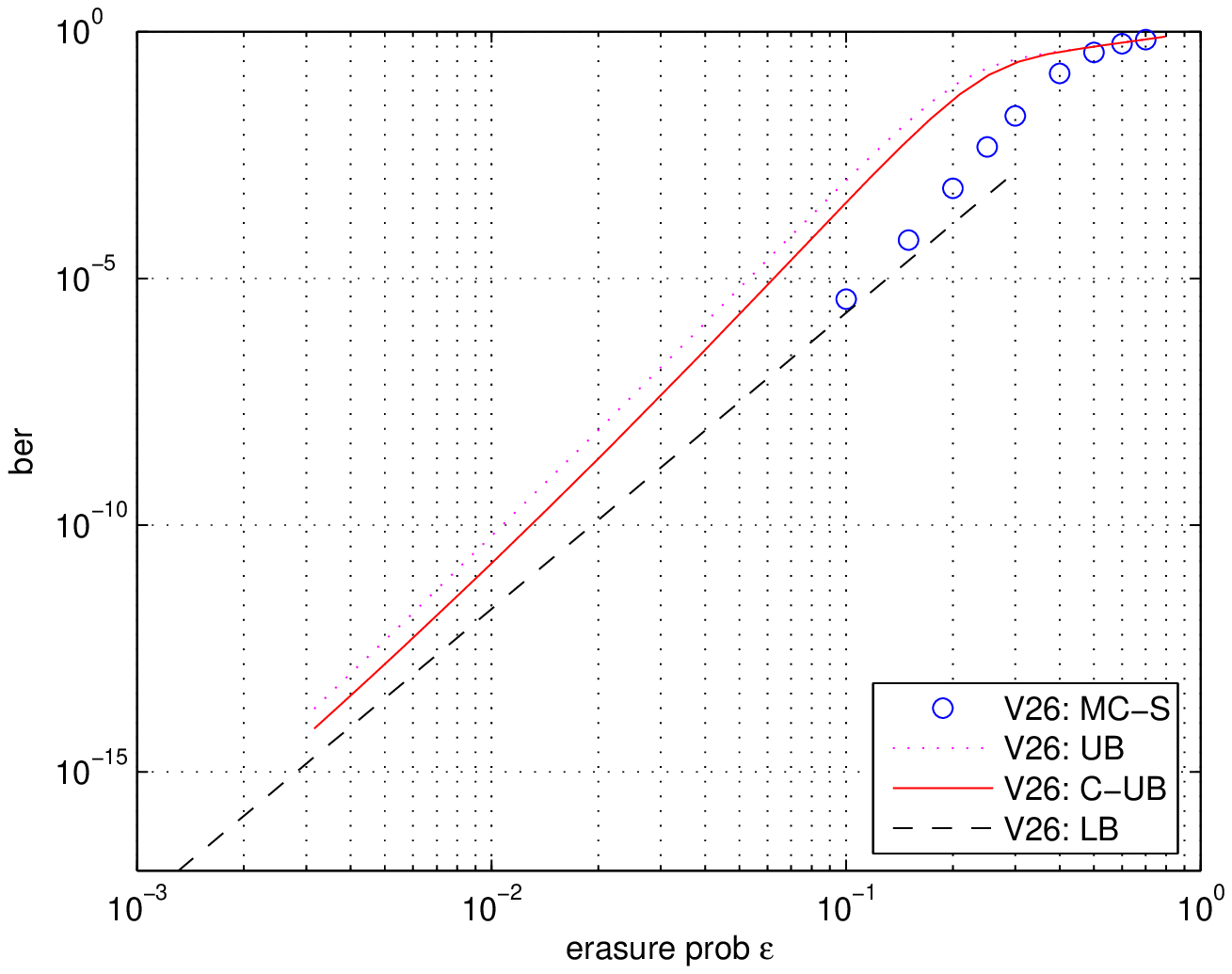}}
&\parbox[\hei]{5.5cm}{\vspace{\mhe}\includegraphics[width=6cm,
keepaspectratio=true]{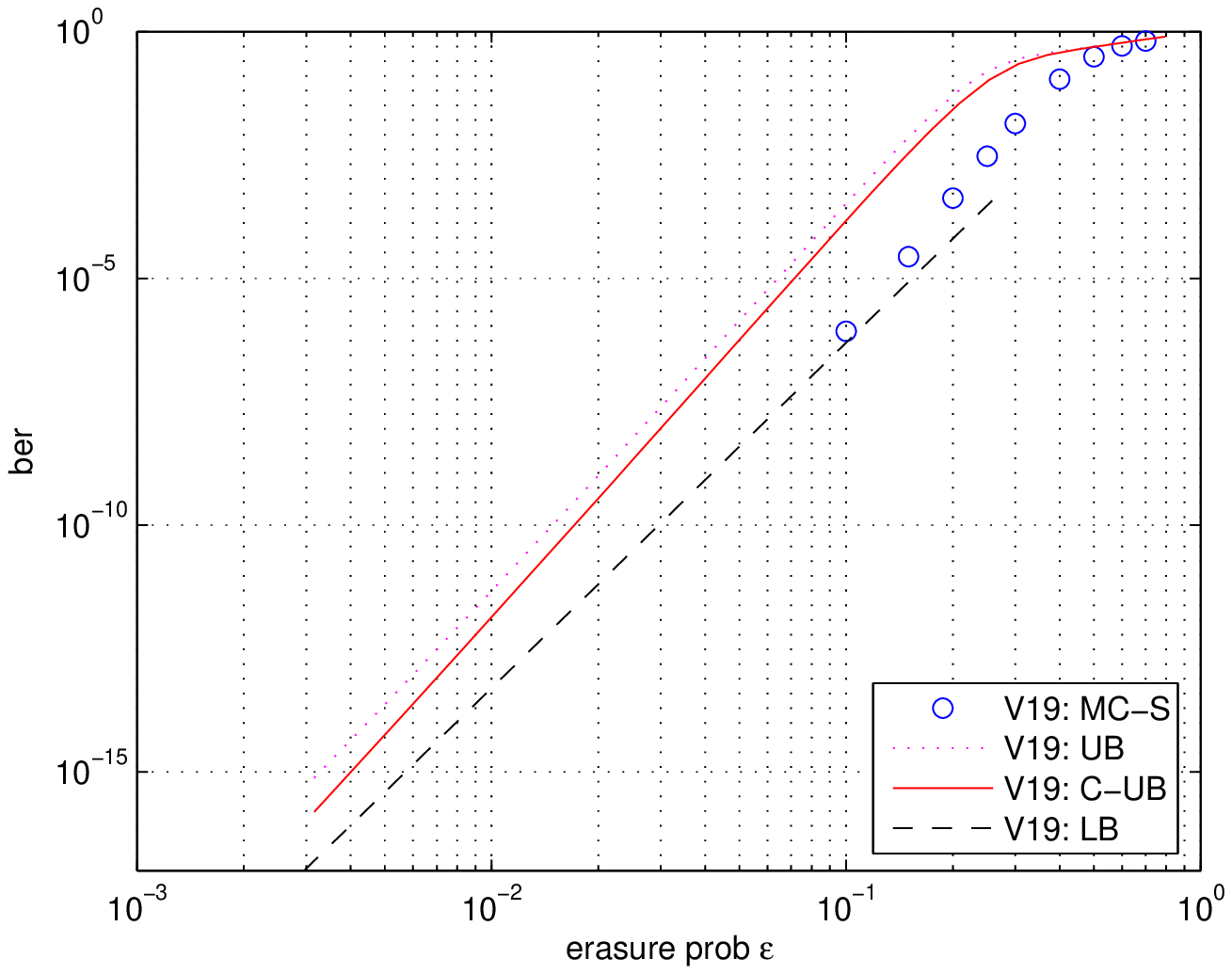}}
\end{tabular}
\caption{Comparisons among the UB, the C-UB, the MC-S, and the LB
for bits~0, 26, and~19 of a randomly generated (3,6) LDPC code
with $n=50$. The asymptotic (order, multiplicity) pairs of the UB,
the C-UB, and the actual BER are V0: $\{(4,1), (4,1), (4,1)\}$,
V26: $\{(6,2), (6,4), (6,2)\}$, and V19: $\{(7,135), (7,10),
(7,5)\}$. \label{fig:R36n50}}
\end{figure*}
\begin{figure*}
\begin{tabular}{ccc}
\parbox[\hei]{5.5cm}{\vspace{\mhe}\includegraphics[width=6cm,
keepaspectratio=true]{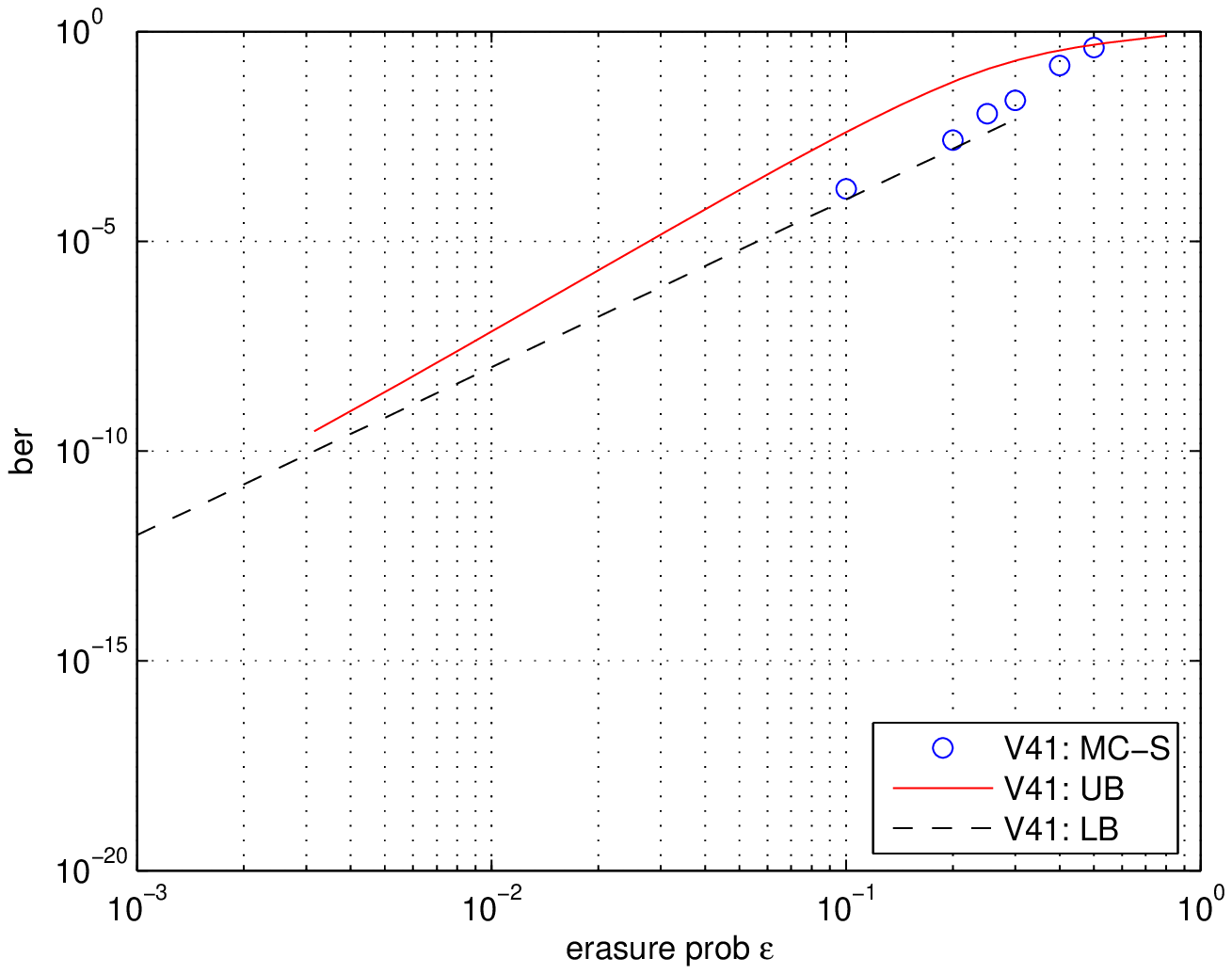}}&\parbox[\hei]{5.5cm}{\vspace{\mhe}\includegraphics[width=6cm,
keepaspectratio=true]{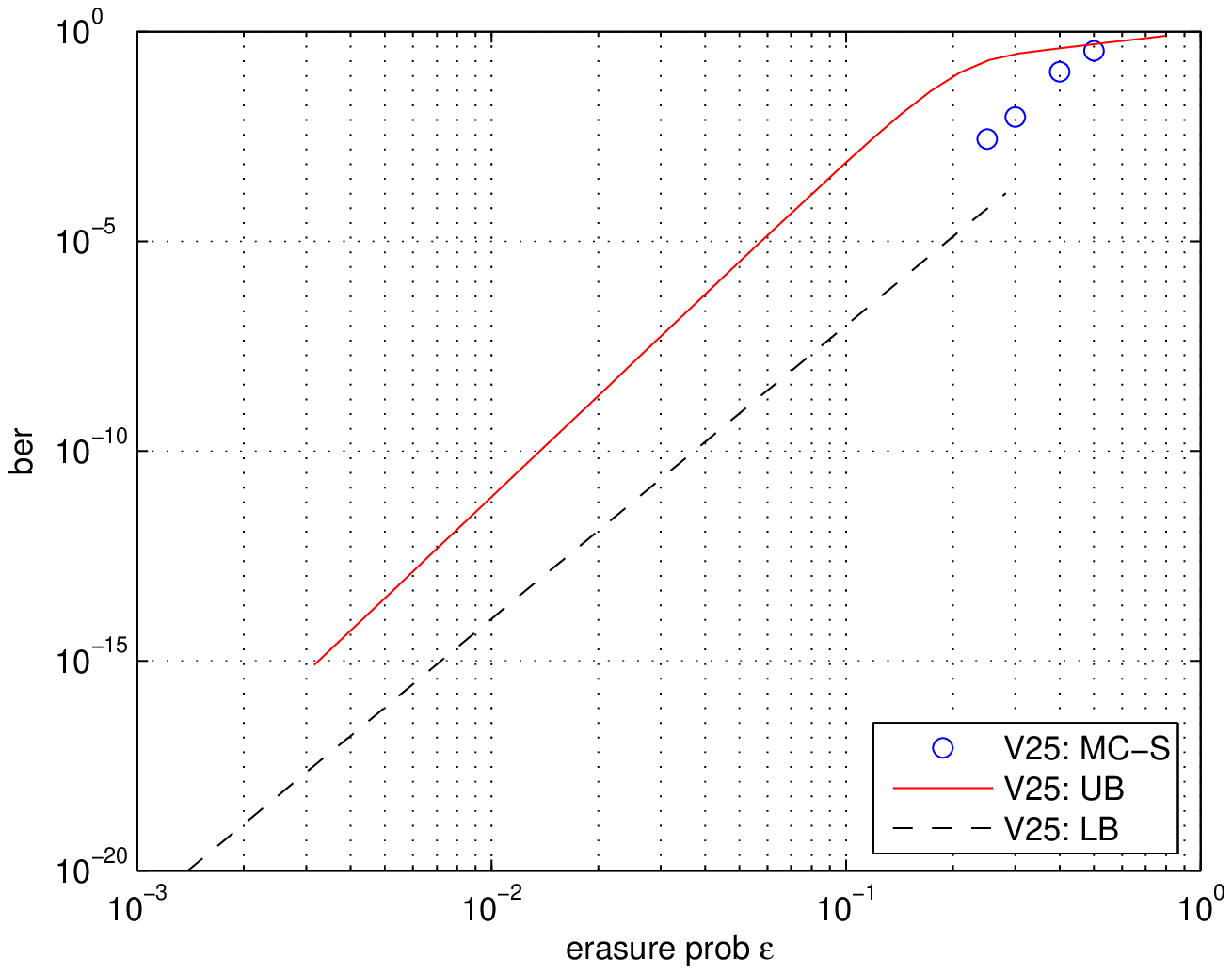}}
&\parbox[\hei]{5.5cm}{\vspace{\mhe}\includegraphics[width=6cm,
keepaspectratio=true]{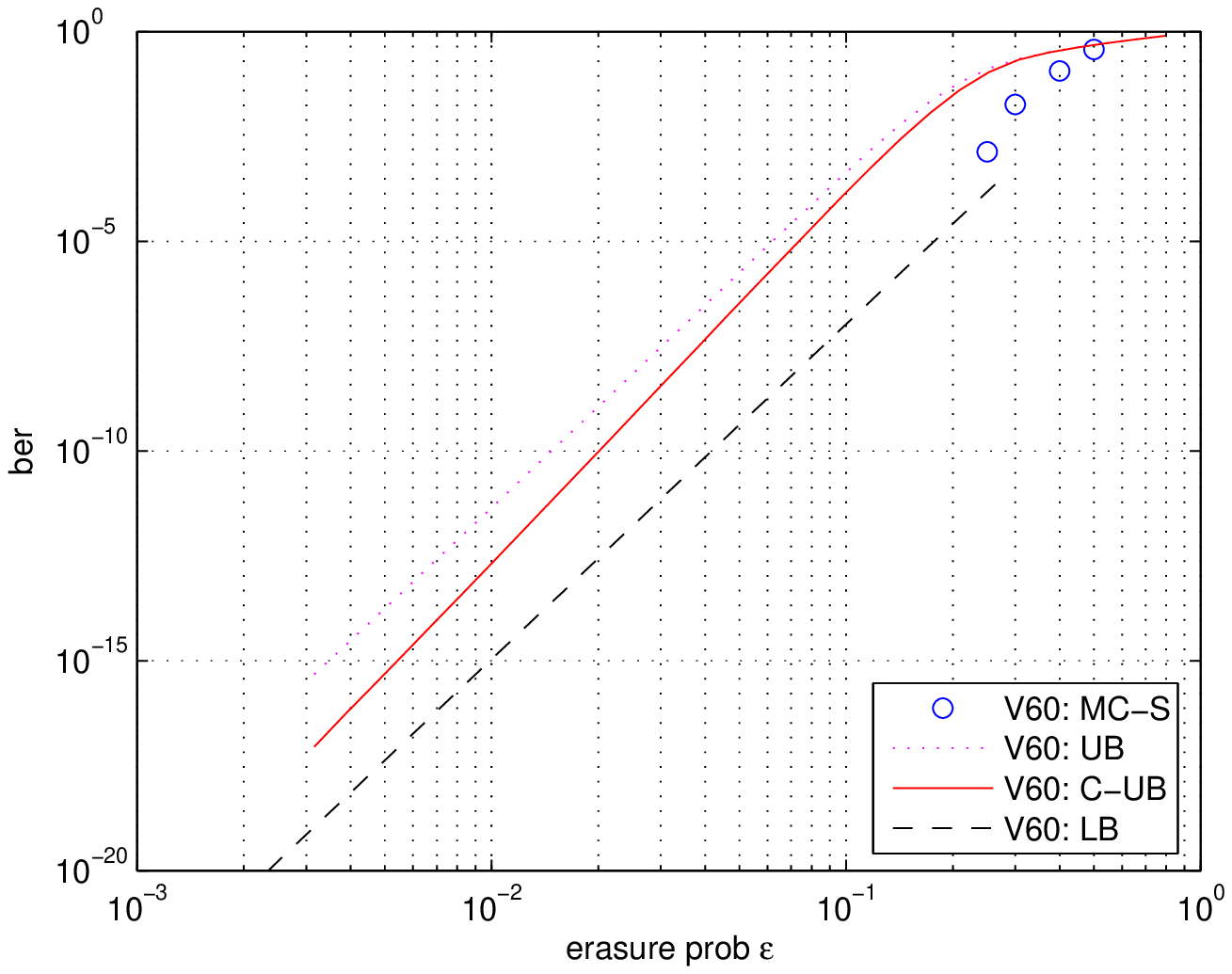}}
\end{tabular}
\caption{Comparisons among the UB, the C-UB, the MC-S, and the LB
for bits~41, 25, and~60 of a randomly generated (3,6) LDPC code
with $n=72$. The asymptotic (order, multiplicity) pairs of the UB,
the C-UB, and the actual BER are V41: $\{(4,1), -, (4,1)\}$, V25:
$\{(7,1), -, (7,1)\}$, and V60: $\{(7,19), (8,431), (8,11)\}$.
\label{fig:R36n72}}
\end{figure*}

\subsection{Performance}

\subsubsection{The (23,12) Binary Golay Code}
The standard parity check matrix of the Golay code is considered.
\Figure{\ref{fig:Golay}} compares the upper bound (UB), the
composite upper bound (C-UB), the Monte-Carlo  simulation (MC-S),
and the side product, the tight lower bound (LB), on bits~0, 5,
and~20. As illustrated, C-UB and LB tightly bracket the MC-S
results, which shows that our UB and C-UB are capable of
decoupling even {\it non-sparse} Tanner graphs with plenty of
cycles.


\subsubsection{A (3,6) LDPC Code with $n=50$}
A (3,6) LDPC code with $n=50$ is randomly generated, and the UB,
the C-UB, the MC-S, and the tight LB are performed on
 bits~0, 26, and~19, as plotted in
 \Figure{\ref{fig:R36n50}}, and the statistics of all 50 bits
are provided in \Table{\ref{tab:perf-stat}(a)}. Our UB is tight in
the asymptotic order for {\it all} bits while 34 bits are tight in
multiplicity. Among the 16 bits not tight in multiplicity, 11 bits
are within a factor of three of
the actual multiplicity. 
In contrast with the Golay code example, the
tight performance can be attributed to the sparse connectivity of
the corresponding Tanner graph. As can be seen in
\Figure{\ref{fig:R36n50}(c)}, the
 C-UB possesses the greatest advantage over those
UBs without tight multiplicity. The C-UB and the LB again tightly
bracket the asymptotic performance.


\subsubsection{A (3,6) LDPC Code with $n=72$}
The UB, the C-UB,  the MC-S, and the tight LB are applied to
 bits~41, 25, and~60, as plotted in
 \Figure{\ref{fig:R36n72}} and the statistics
are in \Table{\ref{tab:perf-stat}(b)}. Almost all asymptotic
orders 
 can be captured by the UB with only two exception bits. Both of
the exception bits are of order 8, which is computed by applying
the C-UB to each bit respectively. 

\subsubsection{(3,6) LDPC Codes with $n=144$}
Complete statistics are presented in
\Table{\ref{tab:perf-stat}(c)}, and we start to see many examples
(101 out of 144 bits) in which our simple UB is not able to
capture the asymptotic order. For those bits, we have to resort to
the C-UB for tighter results.  It is worth noting that the simple
UB is able to identify some bits with order 9, which requires
${144\choose 9}=5.7\times10^{13}$ trials if a brute force method
is employed. Furthermore, {\it all} stopping sets of size $\leq 7$
have been identified, which shows that \Algorithm{\ref{alg:tree1}}
is able to generate tight $\UB$s when only FERs are considered.
Among all our experiments, many of which are not reported herein,
the most computationally friendly case is when considering FERs
for {\it irregular codes} with many degree 2 variable nodes, which
are one of the most important subjects of current research. In
these scenarios, all stopping sets of size $\leq 13$ have been
identified for non-trivial irregular codes with $n=576$, which
evidences the superior efficiency of the proposed algorithm.

\section{Conclusion \& Future Directions}
A new technique upper bounding the BER of any finite code on BECs
has been established, which, to our knowledge, is the first
algorithmic result guaranteeing finite code performance  while
admitting efficient implementation. Preserving much of the
decoding tree structure, this bound corresponds to a narrowing
search of stopping sets. The asymptotic tightness of this
technique has been proved, while the experiments demonstrate the
inherent efficiency of this method for codes of moderate sizes
$n\leq 300$.
One major application of this upper bound is to design high rate
codes with guaranteed asymptotic performance, and our results
specify both the attainable low BER, e.g., $10^{-15}$, and at what
SNR it can be achieved. One further research direction is on
extending the setting to  binary symmetric channels with quantized
belief propagation decoders such as Gallager's decoding
algorithms~A and~B.

\vspace{-.07cm}

%
\bibliography{chihw}

\begin{thebibliography}{1}
\providecommand{\url}[1]{#1}
\csname url@rmstyle\endcsname
\providecommand{\newblock}{\relax}
\providecommand{\bibinfo}[2]{#2}
\providecommand\BIBentrySTDinterwordspacing{\spaceskip=0pt\relax}
\providecommand\BIBentryALTinterwordstretchfactor{4}
\providecommand\BIBentryALTinterwordspacing{\spaceskip=\fontdimen2\font plus
\BIBentryALTinterwordstretchfactor\fontdimen3\font minus
  \fontdimen4\font\relax}
\providecommand\BIBforeignlanguage[2]{{%
\expandafter\ifx\csname l@#1\endcsname\relax
\typeout{** WARNING: IEEEtran.bst: No hyphenation pattern has been}%
\typeout{** loaded for the language `#1'. Using the pattern for}%
\typeout{** the default language instead.}%
\else
\language=\csname l@#1\endcsname
\fi
#2}}

\bibitem{YedidaSudderthBouchaud01}
J.~S. Yedidia, E.~B. Sudderth, and J.-P. Bouchaud, ``Projection algebra
  analysis of error correcting codes,'' Mitsubishi Electric Research
  Laboratories, Technical Report TR2001-35, 2001.

\bibitem{Richardson03}
T.~Richardson, ``Error floors of {LDPC} codes,'' in \emph{Proc. 41st Annual
  Allerton Conf. on Comm., Contr., and Computing}.\hskip 1em plus 0.5em minus
  0.4em\relax Monticello, IL, USA, 2003.

\bibitem{HolzlohnerMahadevanMenyukMorrisZweck05}
R.~{Holzl\"{o}hner}, A.~Mahadevan, C.~R. Menyuk, J.~M. Morris, and J.~Zweck,
  ``Evaluation of the very low {BER} of {FEC} codes using dual adpative
  importance sampling,'' \emph{IEEE Commun. Letters}, vol.~9, no.~2, pp.
  163--165, Feb. 2005.

\bibitem{KrishnanShankar0000}
K.~M. Krishnan and P.~Shankar, ``On the complexity of finding stopping distance
  in {Tanner} graphs,'' preprint.

\bibitem{AmraouiUrbankeMontanariRichardson04}
A.~Amraoui, R.~Urbanke, A.~Montanari, and T.~Richardson, ``Further results on
  finite-length scaling for iteratively decoded {LDPC} ensembles,'' in
  \emph{Proc. IEEE Int'l. Symp. Inform. Theory}.\hskip 1em plus 0.5em minus
  0.4em\relax Chicago, 2004.

\bibitem{DiProiettiTelatarRichardsonUrbanke02}
C.~Di, D.~Proietti, E.~Telatar, T.~J. Richardson, and R.~L. Urbanke,
  ``Finite-length analysis of low-density parity-check codes on the binary
  erasure channel,'' \emph{IEEE Trans. Inform. Theory}, vol.~48, no.~6, pp.
  1570--1579, June 2002.

\bibitem{StepanovChernyakChertkovVasic0000}
M.~G. Stepanov, V.~Chernyak, M.~Chertkov, and B.~Vasic, ``Diagnosis of
  weaknesses in modern error correction codes: a physics approach,''
  \emph{Phys. Rev. Lett.}, to be published.

\end{thebibliography}
\bibliographystyle{IEEEtran}

\end{document}